\documentclass[doublecol]{epl2}
\usepackage{amsmath}
\usepackage{amsfonts}

\newcommand{\kronecker}{\raisebox{1pt}{\ensuremath{\:\otimes\:}}}
\newcommand{\ltsimeq}{\raisebox{-0.6ex}{$\,\stackrel
        {\raisebox{-.2ex}{$\textstyle <$}}{\sim}\,$}}

\title{Analysis of parameter mismatches in the master stability function for network synchronization}
\shorttitle{Analysis of parameter mismatches in the MSF for network synchronization} 

\author{Francesco Sorrentino\inst{1}, Maurizio Porfiri\inst{2} }
\shortauthor{F. Sorrentino and M. Porfiri}

\institute{
  \inst{1} Universit{\`a} degli Studi di Napoli Parthenope, 80143 Napoli, Italy\\
  \inst{2} Polytechnic Institute of New York University, Brooklyn, NY 11201, USA
}

\abstract{
In this letter, we perform a sensitivity analysis on the  master stability function approach for the synchronization of networks of coupled dynamical systems. More specifically, we analyze the linear stability of a nearly synchronized solution for a network of coupled dynamical systems, for which the individual dynamics and output functions of each unit are approximately identical and the sums of the entries in the rows of the coupling matrix slightly deviate from zero. The motivation for this parametric study comes from experimental instances of synchronization in human-made or natural settings, where ideal conditions are difficult to observe.}

\begin{document}

\maketitle

\textbf{Introduction.} Synchronization of networks of coupled dynamical systems has been the subject of intensive research, see for example the reviews \cite{Report2,SUCNS,Sreport,Miranda}. Chaos synchronization of networked dynamical systems finds applications in secure communication \cite{Cu:Op,Argyris,Feki}, system identification \cite{Abarbanel, Abarbanel2,Abarbanel4,IDTOUT}, data assimilation \cite{So:Ott:Day,Duane}, sensors \cite{SOTT2}, 
information encoding and transmission \cite{Ha:Gr:Ott,Dr:He:An:Ott}, and multiplexing \cite{Ts:Su}. In this framework, the master stability function analysis provides a necessary and sufficient condition for the linear stability of the synchronous solution.
However, most of the research on this approach focuses on ideal conditions, which are difficult to implement in experiments. 

We consider a typical experimental scenario for a set of dynamical systems 
that are coupled through a network to achieve synchronization. We assume that each of the elements which constitute the experiment is selected to reflect certain nominal characteristics; yet, we allow these components to be affected by small mismatches from their nominal values. We consider a wide range of possible deviations from nominal operating conditions that may affect simultaneously the individual units' dynamics, the individual units' output functions, and the coupling gains among the systems.
Another motivation for the proposed analysis is the study of the collective behavior of biological groups, where individuals are generally different in nature and their couplings are typically affected by fluctuations about an average  or \emph{nominal} value; see for example \cite{Sumpter}. 

We consider the following equations of motion for a set of coupled chaotic systems in their nominal conditions
\begin{equation}
\dot{x}_i(t)=F(x_i(t))+ \sigma \sum_{j=1}^{N} A^{NOM}_{ij} H(x_j(t)), \quad i=1,2,...,N, \label{NP}
\end{equation}
where ${x}_i \in {\mathbb{R}}^n$ is the $n$-dimensional vector
describing the state of  node $i$, $F:
{\mathbb{R}}^n \to {\mathbb{R}}^n$ governs the uncoupled dynamics of node $i$, $H: {\mathbb{R}}^n \to {\mathbb{R}}^n$ is a vectorial
output function, $\sigma$ is a 
scalar gain describing the overall coupling
strength, and $N$ is the number of nodes in the network. The network is defined by the matrix $A^{NOM}=\{A^{NOM}_{ij} \}$, describing the 
coupling from node $j$ to node $i$. We refer to equation set (\ref{NP}) as nominal, as we assume that it corresponds to a given experimental design. 
A sufficient condition for the existence of a synchronized solution,
\begin{equation}
x_1(t)=x_2(t)=...=x_N(t)=x_s(t),\label{ss}
\end{equation}
is that
\begin{equation}
\sum_{j} A^{NOM}_{ij}=0, \quad i=1,...,N, \label{condition}
\end{equation}
that is, all the row-sums\footnote{In what follows, we sometimes refer to the \emph{row-sums} of a matrix, indicating with this terminology the sums of the entries along the rows of the matrix. We further comment that the
analysis stays unaltered if the right hand side of (\ref{ss}) equals a constant.}
 of the matrix $A^{NOM}$ are equal to zero.
In case  condition (\ref{condition}) is satisfied, a synchronized solution $x_s(t)$ exists that satisfies
\begin{equation}
\dot{x}_s(t)=F(x_s(t)).
\end{equation}
 We use $\ell^{NOM}_1,...,\ell^{NOM}_N$ to identify the eigenvalues of the matrix $A^{NOM}$, which are in general complex numbers. Note that (\ref{condition}) implies that $A^{NOM}$ has one eigenvalue, $\ell^{NOM}_1=0$, with associated right eigenvector $1_N=[1,1,...,1]$.

 The linear stability of (\ref{ss}) can be assessed by using the master stability function \cite{FujiYama83,Pe:Ca}. Within this framework, the synchronous solution is stable if the maximum Lyapunov exponent associated with the parametric equation
\begin{equation}\label{c}
\dot{\gamma}(t)=[DF(x_s(t))+c DH(x_s(t))] \gamma(t)
\end{equation}
is negative for every $c=\sigma \ell^{NOM}_k$,  $k=2,...,N$, where $\gamma$ is an $n$-dimensional vector. Then it is possible to associate a master stability function to Eq. (\ref{c}), which yields the maximum Lyapunov exponent of  (\ref{c}) as a function of the parameter $c$. 
Thus stability of the synchronized solution can be assessed for any given network described by Eqs. (\ref{NP})  by verifying that the corresponding eigenvalues $\sigma \ell^{NOM}_k$,  $k=2,...,N$, are within the region of the complex plane for which the master stability function is negative.  

\textbf{Problem statement.} The assumptions underlying the set of equations (\ref{NP}) are that:

 (i) the individual units are all described by the same dynamics $\dot{x}_i(t)=F(x_i(t))$;

   (ii) the systems' outputs are all described by the same function $H$;

   (iii) the sums of the rows of the matrix $A^{NOM}$ are all zero,
that is, condition (\ref{condition}) is verified at each node $i=1,...,N$.

While assumptions (i)-(iii) can be easily reproduced in a numerical simulation, their practical implementation in experiments is challenging. Qualitatively good satisfaction of (i)-(iii) in experimental instances of synchronization often requires fine tuning \cite{Fi:Jo:Ca,EXPS1,EXPS2,EXPS3,EXPS4,EXPS5,EXP,EXP2}. In \cite{SOTT,EXP,EXP2}, an adaptive strategy  to dynamically preserve synchronization in the presence of slow a-priori-unknown time-variations of the couplings is proposed. Though such strategy is able to preserve condition (iii) in the presence of external perturbations,   the row-sums of the coupling matrix are  typically non zero over the time scale of the adaptation.

{In [30,31],  assumption (i) is removed and the effect of small mismatch of the individual units is considered. That is, these works consider the case where $F$ in Eq. (1) is replaced by $F_i$ and the difference between $F$ and $F_i$ is small. In this letter, we extend the considerations of [30,31] to simultaneously allow for deviations from the exact satisfaction of all three of the assumptions (i), (ii), and (iii).}  Namely,
we assume that (i),(ii), and (iii) are nominal design conditions, which might not be exactly reproduced in an experiment. 
We show that if all the mismatches are small as compared to the nominal conditions, the linear stability of the nearly synchronized solution can be studied by using an extended master stability function. {Moreover, when the nearly synchronous evolution is stable, the mismatches introduce forcing terms in the parametric equation that maintain the network in a state of approximate synchronization.}
 
To take into account approximate, rather than exact satisfaction of (i), (ii), and (iii), we rewrite the network equations in the form
\begin{equation}
\dot{x}_i(t)=F(x_i(t), m_i)+ \sigma \sum_{j=1}^{N} A_{ij} H(x_j(t), p_j),  \label{NP2}
\end{equation}
$i=1,2,...,N$, where  $A_{ij}$ represents the coupling from node $j$ to node $i$, $m_i$ is a parameter used to identify variations of the dynamics at each node $i$, and $p_i$ is a parameter of the output function of each node $i$. We assume that $m_i=\bar{m}+\delta m_i$, where $\bar{m}={N}^{-1}\sum_i m_i$ and $\delta m_i$ is a small mismatch. Similarly, we write $p_i=\bar{p}+\delta p_i$, where $\bar{p}={N}^{-1}\sum_i p_i$ and $\delta p_i$ is a small mismatch. Note that by construction $\sum_i \delta m_i=0$ and $\sum_i \delta p_i=0$. The elements $A_{ij}$'s represent imperfect realizations of the nominal couplings $A_{ij}^{NOM}$'s, that is, $A_{ij}=A_{ij}^{NOM}+\delta A_{ij}$, $i,j=1,...,N$, where  $\delta A_{ij}$ is a small mismatch. In general, in the presence of deviations of the $A_{ij}$'s from their nominal values, it is not possible to write a condition equivalent to (\ref{condition})  and thus to extend directly the master stability function formalism. 
For small $\delta A_{ij}$'s, we can write 
\begin{equation}
\sum_{j} A_{ij}= \sum_{j} \delta A_{ij}=\delta \bar{a}+\delta a_i, \label{cond2}
\end{equation}
where
\begin{equation}
\delta \bar{a}=N^{-1} \sum_{i,j} A_{ij}=N^{-1} \sum_{i,j} \delta A_{ij}
\end{equation}
is the average sum of the rows of the matrix $A$ 
and,
\begin{equation}\label{dai}
\delta a_i=\Bigl( \sum_{j} \delta A_{ij} \Bigr) -\delta \bar{a}=\Bigl( \sum_{j} \delta A_{ij} \Bigr) - N^{-1} \sum_{ij} \delta A_{ij}
\end{equation}
is a small deviation. The deviations $\delta a_i$ are calculated with respect to the average row-sum $\delta \bar{a}$, hence they have zero sum, that is, $\sum_j \delta a_j=0$.
By using condition (\ref{cond2}) in equation set (\ref{NP2}), we obtain 
\begin{equation}\begin{split}
\dot{x}_i(t)= & F(x_i(t), m_i)+ \sigma \sum_{j} A'_{ij} H(x_j(t), p_j)+ \\
 & \sigma \delta a_i H(x_i(t),p_i),  \label{NU}
\end{split}\end{equation}
$i=1,2,...,N$, where we have introduced the matrix $A'$ defined by
\begin{align} \label{A}
A'_{ij}= \left\{ \begin{array} {ccc} {A_{ij},} \quad & \mbox{if} \quad {j \neq i,} \\ {A_{ii}-\delta a_i,}  \quad & \mbox{if} \quad    {j=i.} \end{array} \right. 
\end{align}
By construction, the matrix $A'=\{A'_{ij}\}$ is such that the sums of its rows are constant and equal to $\delta \bar{a}$. We note that by setting to zero all the mismatches $\delta m_i$, $\delta p_i$, and $\delta a_i$ in (\ref{NU}),  a synchronized solution exists for the set of equations in (\ref{NU}) of the form
\begin{equation}\label{sync}
\dot{\tilde{x}}_s=F({\tilde{x}}_s,\bar{m})+\sigma \delta \bar{a} H({\tilde{x}_s},\bar{p}).
\end{equation}

{\bf Extended master stability function.} We introduce the average trajectory $\bar{x}(t)=N^{-1} \sum_{k} x_k(t)$ that satisfies the following average dynamics
\begin{equation}\label{preexp2}
\begin{split}
\dot{\bar{x}}(t)=  N^{-1}  \Bigl[  \sum_k F(x_k(t),m_k) +  \sigma \sum_{k,j} A'_{kj} H(x_j(t),p_j)+ \\
 \sigma \sum_k \delta a_k H(x_k(t),p_k) \Bigr].
\end{split}
\end{equation}
Since the quantities $\delta a_i$,  $\delta p_i$, and $\delta m_i$ are small, we expect the individual trajectories $x_i(t)$ to be \emph{close} to the average trajectory $\bar{x}(t)$, that is, $\|x_i(t)-\bar{x}(t)\| \leq k^*$ for all times and  some small $k^*>0$. We define the variation with respect to the average trajectories as $\delta x_i(t)=(x_i(t)-\bar{x}(t))$. By expanding both (\ref{NP2}) and (\ref{preexp2}) to first order about $(\bar{x}(t),\bar{m},\bar{p})$, we obtain
\begin{equation}
\begin{split}
\delta \dot{x}_i(t)=& DF_x(\bar{x}(t),\bar{m}) \delta x_i(t)+ DF_{m}(\bar{x}(t),\bar{m}) \delta m_i +    \\
& \sigma DH_x(\bar{x}(t),\bar{p}) \sum_j  (A'_{ij}-b_j)  \delta x_j(t) +\\ & \sigma DH_p(\bar{x}(t),\bar{p}) \sum_j  (A'_{ij}-b_j)  \delta p_j+  \\ & \sigma H(\bar{x}(t),\bar{p}) \delta a_i,\label{postexp2}
\end{split}
\end{equation}
$i=1,...N$, where $b_j=N^{-1} \sum_k A'_{kj}$, that is, $b_j$ represents the sum of the entries over  column $j$ of the matrix $A'$ divided by $N$, for $j=1,...,N$. We have indicated with $DF_x$  and $DH_x$ the partial derivative of the functions $F$ and $H$ with respect to $x$, with  $DF_m$ the partial derivative of the function $F$ with respect to $m$, and with  $DH_p$ the partial derivative of the function $H$ with respect to $p$.  To obtain (\ref{postexp2}), we have used the properties  $\sum_j \delta x_j=0$, $\sum_j \delta a_j=0$, $\sum_j \delta m_j=0$, $\sum_j \delta p_j=0$, $\sum_j b_j=\delta \bar{a}$, and we have discarded second order terms in all the variations.

 We define $\ell'_1,\ell'_2,...,\ell'_N$ as the eigenvalues of the matrix $A'$. We note that since the row-sums of the matrix $A'$ are equal to $\delta \bar{a}$, the matrix $A'$ has one eigenvalue $\ell'_1=\delta \bar{a}$, with associated right eigenvector $1_N$.
Now, we consider the matrix $\tilde{A}=\{\tilde{A}_{ij}\}$, where $\tilde{A}_{ij}=(A'_{ij}-b_j)$, and we look for the solutions of the eigenvalue equation $\tilde{A}\bar{v}_i=\lambda_i \bar{v}_i$. We observe that the matrix $\tilde{A}$ has the property that both the sums of its rows and its columns are equal zero. Thus $\bar{v}_1=1_N$ is still a right eigenvector for the matrix $\tilde{A}$, with associated eigenvalue $\lambda_1=0$. 
Moreover, $\bar{w}_1=1_N$ is also the left eigenvector of the matrix $\tilde{A}$, associated with the eigenvalue $0$. 
The remaining eigenvalues of the matrix $\tilde{A}$ are $\lambda_i=\ell'_i$ for $i=2,...,N$ \cite{Su:Bo:Ni}.  In other words, the matrices $A'$ and $\tilde{A}$ have the same spectrum except for the eigenvalue associated with the right eigenvector $\bar{v}_1=1_N$. As discussed in what follows, the eigenvalues $\lambda_2,...,\lambda_N$ control the stability of the nearly-synchronous solution.

Equations (\ref{postexp2}) can be rewritten as
\begin{equation}\begin{split}
\delta \dot{X}(t)= & [ I_N \kronecker DF_x(\bar{x}(t),\bar{m}) + \sigma \tilde{A} \kronecker DH_x(\bar{x}(t),\bar{p}) ] \delta {X}(t)+ \\  & [I_N \kronecker DF_{m}(\bar{x}(t),\bar{m})] \delta {M}+ \\  & \sigma [\tilde{A} \kronecker DH_{p}(\bar{x}(t),\bar{p})] \delta {P} + \\ & \sigma [I_N \kronecker H(\bar{x}(t),\bar{p})] \delta {A},\label{kron2}
\end{split}\end{equation}
where $\delta {X}(t)=[\delta {x}_1(t)^T,\delta {x}_2(t)^T,...,\delta {x}_N(t)^T]^T$, $\delta {M}=[\delta {m}_1,\delta {m}_2,...,\delta {m}_N]^T$, $\delta {P}=[\delta {p}_1,\delta {p}_2,...,\delta {p}_N]^T$, $\delta {A}=[\delta {a}_1,\delta {a}_2,...,\delta {a}_N]^T$, and the symbol $\kronecker$ indicates direct product or Kronecker product.

Following \cite{Pe:Ca} and assuming that the matrix $\tilde{A}$ is diagonalizable, we write, $\tilde{A}=V \Lambda W$, where $\Lambda=\mbox{diag}(\lambda_1,\lambda_2,...,\lambda_N)$,  $V$ is a matrix whose columns are the right eigenvectors of the matrix $\tilde{A}$, and $W=V^{-1}$. Premultiplying (\ref{kron2}) by $W \kronecker I_n$, we obtain
\begin{equation} \begin{split}
 \dot{Q}(t)=& [ I_N \kronecker DF_x(\bar{x}(t),\bar{m}) + \sigma \Lambda \kronecker DH_x(\bar{x}(t),\bar{p}) ]  {Q}(t) +\\  & [W \kronecker DF_{m}(\bar{x}(t),\bar{m})] \delta {M}+\\ & \sigma [\lambda_i W \kronecker DH_{p}(\bar{x}(t),\bar{p})] \delta {P} +\\  & \sigma [W \kronecker H(\bar{x}(t),\bar{p})] \delta {A},\label{q}
\end{split} \end{equation}
where  $Q(t)=(W \kronecker I_n) \delta X(t)$. We note that both matrices $I_N$ and $\Lambda$ in the homogeneous part of Eq. (\ref{q}) are diagonal matrices. Thus equation (\ref{q}) can be decomposed into $N$ blocks of the form
\begin{equation} \begin{split}
\dot{ q}_i(t)=& [ DF_x(\bar{x}(t),\bar{m}) + \sigma \lambda_i DH_x(\bar{x}(t),\bar{p}) ] {q}_i(t)+   \\  & \sum_j W_{ij} \delta {m_j} DF_{m}(\bar{x}(t),\bar{m}) + \\  & \sigma \lambda_i  \sum_j W_{ij} \delta {p_j} DH_p(\bar{x}(t),\bar{p})+ \\ & \sigma \sum_j W_{ij} \delta {a_j} H(\bar{x}(t)),\label{blocks}
\end{split} \end{equation}
$i=1,...,N$. We comment that the homogeneous part of each block in (\ref{blocks}) is independent of the other blocks. For $i=1$, the variational equation (\ref{blocks}) yields $q_1(t)=0$ since $\sum_{i=1}^{N}\delta x_i(t)=0$ and $1_N$ is a left eigenvector.
Thus we note that the component of the evolution along the direction $x_1=x_2=...=x_N$ is not affected by the mismatches $\delta m_i$, $\delta p_i$, and $\delta a_i$. Stability of the nearly-synchronized solution is controlled by perturbations in the remaining directions, $q_2,...,q_N$. Following \cite{Pe:Ca,Su:Bo:Ni}, it is possible to associate  the following parametric equation to Eq. (\ref{blocks})
\begin{equation}\begin{split}
\dot{ z}(t)= & [ DF_x(\bar{x}(t),\bar{m}) + \omega DH_x(\bar{x}(t),\bar{p}) ] {z}(t)  + \\  & \epsilon DF_{m}(\bar{x}(t),\bar{m}) + \zeta DH_p(\bar{x}(t),\bar{p})+\eta H(\bar{x}(t)),\label{LD}
\end{split}\end{equation}
which corresponds to equation set (\ref{blocks}) upon setting  $z=q_i$, $\omega=\sigma \lambda_i$,  $\epsilon=\sum_j W_{ij} \delta {m_j}$, $\zeta=\sigma \lambda_i  \sum_j W_{ij} \delta {p_j}$, and $\eta=\sigma \sum_j W_{ij} \delta {a_j}$, for $i=2,...,N$.

In order to assess the linear stability of the nearly-synchronous solution, Eq. (\ref{LD}) needs to be tested for the set of eigenvalues $\lambda_2,...,\lambda_N$. 
If the Lyapunov exponents associated with the homogeneous part of Eq. (\ref{LD}), $i=2,...,N$ are negative, the nearly-synchronous solution is stable. In this case, the forcing terms on the right hand side of Eq. (\ref{LD}), $i=2,...,N$, can be considered as inputs to a stable system. 
 It is then possible to associate an extended  master stability function $\mathcal{M}(\omega,\epsilon,\zeta,\eta)$, defined as $\lim_{\tau \rightarrow \infty} \sqrt{\tau^{-1} \int_0^{\tau} \|z(t) \|^2 dt}$ to Eq. (\ref{LD}), which yields the asymptotic norm of the time average of  $z$ as a function of the tuple  $(\omega,\epsilon,\zeta,\eta)$. However, stability of the nearly-synchronous solution depends on the homogeneous part of (\ref{LD}), that is,  it depends on $\omega$, while it is independent of $\epsilon,\zeta$, and $\eta$. We note that for $\delta a_i=0$, $\delta m_i=0$, and $\delta p_i=0$ with $i=1,...,N$, the parametric  equation (\ref{LD}) reduces to (\ref{c}), which corresponds to the ideal case where all the parameters are equal to their nominal values.

Moreover, following \cite{Su:Bo:Ni}, in the case that the master stability function is  asymptotically bounded and $\omega$ is fixed, we have that $\mathcal{M}(\omega,\epsilon,\zeta,\eta)$ scales linearly with respect to $\epsilon$, $\zeta$, and $\eta$, that is,
\begin{equation}\label{ccc}
\mathcal{M}(\omega,\epsilon,\zeta,\eta) \simeq c_\epsilon(\omega) |\epsilon|+ c_\zeta(\omega) |\zeta|+c_{\eta}(\omega) |\eta|,
\end{equation}
where the coefficients $c_\epsilon,c_\eta$, and $c_\zeta$ are functions of $\omega$.

We comment that the extended master stability function depends on the eigenvalues 
of the \emph{perturbed} matrix $A'$ and not on those of the \emph{nominal} matrix $A^{NOM}$. 
The matrix $A'$ can be considered  a perturbed version of the nominal matrix $A^{NOM}$, $A'=A^{NOM}+\Delta$, where the perturbation matrix $\Delta=\{\Delta_{ij} \}=\{\delta A_{ij} - \delta^{ij} ( \sum_{j} \delta A_{ij} -a)\}$ and $\delta^{ij}$ indicates the Kronecker delta, ${i,j}=1,...,N$. Note the sums of the rows of $\Delta$ are equal to $\delta \bar{a}$. The eigenvalues of the perturbed matrix $A'$ can be computed from the spectral properties of  $A^{NOM}$. By using classical perturbation theory \cite{Matrix}, and assuming that the eigenvalues of the matrix $A^{NOM}$ are all distinct, we find
\begin{equation} \label{pert}
\lambda_i \simeq \ell_i^{NOM}+\frac{{\hat{w}_i}^T \Delta  \hat{v}_i}{\hat{w}_i^T \hat{v}_i}, \qquad i=2,...,N,
\end{equation}
where $\hat{w}_i$ and $\hat{v}_i$ are the left and right eigenvectors associated with the eigenvalues $\ell_i^{NOM}$ of the matrix $A^{NOM}$, respectively.
Equation (\ref{pert}) shows that the deviations of the relevant eigenvalues from their nominal values are on the same order of the perturbations $\Delta_{ij}$ on the couplings. We also comment that Eq. (\ref{pert}) predicts that $\ell'_1 \simeq ({{\hat{w}_1}^T \Delta  1_N})/({{\hat{w}_1}^T 1_N})=a$, since  $\ell'_1=a$ by construction. Similar arguments can be used to estimate the left eigenvectors of  $\tilde{A}$ from the spectral properties of $A^{NOM}$.

The main result of our analysis is that stability of the nearly-synchronous evolution for the system (\ref{NP2}) can be assessed by using a master stability function, which depends on the eigenvalues of an appropriately modified coupling matrix $A'$. Though in a practical situation it is not feasible to exactly calculate these eigenvalues, for small deviations of the couplings from their nominal values they differ from their nominal values $\ell_i^{NOM}$ by a small quantity of the same order of the $\Delta$. Moreover, 
the mismatches in the individual functions $F$ and $H$, along with the deviations in the row-sums of the coupling matrix $A$, introduce forcing terms in Eq. (\ref{LD}) through the coefficients $\epsilon,\zeta$, and $\eta$. Such forcing terms maintain the network in a state of approximate synchronization.

Following \cite{Su:Bo:Ni}, in  case the matrix $\tilde{A}$ has an orthonormal basis of eigenvectors, that is, it is symmetric, we can write
\begin{equation}\label{E}
E \equiv  \lim_{\tau \rightarrow \infty} {\tau}^{-1} \int_0^{\tau} \sum_{i}^{N} \|\delta x_i(t) \|^2 dt=\sum_{i=2}^{N} \mathcal{M}^2(\omega_i,\epsilon_i,\zeta_i,\eta_i).
\end{equation}
We note that $E$ is a quantity of physical interest, as it represents the time average sum, over all the coupled systems of the  \emph{distances} $\| \delta x_i(t) \|$ from the average trajectory $\bar{x}(t)$. One of the advantages of this approach is that, by computing  the master stability function once,  $E$ can be estimated \emph{for any network topology} that approximately satisfies the constant-row-sum condition.

As pointed out in \cite{Su:Bo:Ni}, a complication with this approach is that Eq. (\ref{LD}) depends on $\bar{x}(t)$, which is an averaged trajectory over all the systems in the network. In a large network, calculating $\bar{x}(t)$ may be computationally expensive, as it requires full integration of $N$ individual systems, see Eq. (\ref{preexp2}). However, for practical purposes, $\bar{x}(t)$ in (\ref{LD}) can be replaced by the individual dynamics $\tilde{x}_s(t)$ in (\ref{sync}), 
which depends explicitly on $\bar{m},\bar{p}$, and $\delta \bar{a}$. We comment that, unless precise knowledge of the characteristics of all the individual units and of their couplings is available, it is difficult to exactly compute $\bar{m},\bar{p}$, and $\delta \bar{a}$. Nevertheless, a priori knowledge on the statistical properties of the coupled systems can be used to infer the average parameters. For example, if $m_i$, $p_i$ and $ a_i$, with $i=1,...,N$ are taken as independent and identically distributed  random variables, drawn from distributions having mean corresponding to their nominal values, and finite variance, the central limit theorem states that $\bar{m},\bar{p}$, and $\delta \bar{a}$ approach their nominal values as the number of nodes increases.

{\bf Numerical simulation.}
We use the algorithm in \cite{korea} to generate a scale-free network of $N=100$ nodes with average degree equal to $30$ and exponent of the power-law degree distribution equal to $3$. For each pair of nodes $i,j=1,...,N$, $j \neq i$, $A^{NOM}_{ij}=A^{NOM}_{ji}=1$ if nodes $i$ and $j$ are connected; otherwise, $A^{NOM}_{ij}=A^{NOM}_{ji}=0$. We set $A^{NOM}_{ii}=-\sum_j A^{NOM}_{ij}$, which guarantees that the row-sums of the matrix $A^{NOM}$ are equal to zero. Moreover, as the matrix $A^{NOM}$ is symmetric, it is diagonalizable, its eigenvalues are real, and the eigenvectors can be taken to be orthonormal.   We find that $\ell_2=12.7018$ and $\ell_N=86.0531$, where we set $\ell_1 \leq \ell_2,..., \leq \ell_N$.

We consider that $A_{ij}=A^{NOM}_{ij}(1+\varsigma_a \rho_{ij})$, for $i,j=1,...,N$, where $\rho_{ij}=\rho_{ji}$ is a random number drawn from a standard normal distribution and $\varsigma_a$ is a scalar. Upon this selection, the matrix $A'$ in (\ref{A})  is symmetric; the matrix $\tilde{A}$ 
is also symmetric, since $b_j=\delta \bar{a}/N$ for $j=1,...,N$.  For $\varsigma_a=10^{-4}$, we obtain $\lambda_2=12.7007$ and $\lambda_N=86.0529$. This is in agreement with Eq. (\ref{pert}), as we find that $|\lambda_i-\ell_i|$ is on average on the same order of magnitude of the deviations on the couplings\footnote{We have also performed numerical experiments for $\rho_{ij} \neq \rho_{ji}$ and we have found that, for small values of $\varsigma_a$, the eigenvalues $\lambda_i$'s are still real.}. 

We perform a numerical experiment for a set of nominally identical R\"ossler oscillators that are affected by mismatches in both their dynamics and output functions and are coupled by the scale free network, described by the matrix $A$. In this case, the equations of motion are
\begin{equation}\label{ross}
\begin{split}
\dot{x}_{i1}(t)= & -x_{i2}(t)-x_{i3}(t)+ \sigma \sum_j A_{ij} (x_{j1}(t)+p_j),\\
\dot{x}_{i2}(t)= & x_{i1}(t)+ m_i x_{i2}(t),\\
\dot{x}_{i3}(t)= & 0.2+(x_{i1}(t)-7)x_{i3}(t),
\end{split}
\end{equation}
$i=1,...,N$, where the state vector of oscillator $i$ is $x_i=[x_{i1},x_{i2},x_{i3}]^T$. The parameters $p_j$ are random numbers drawn from a Gaussian distribution with mean zero and standard deviation $\varsigma_p$, and the parameters $m_i$ are random numbers drawn from a Gaussian distribution with mean value equal to $0.2$ and standard deviation $\varsigma_m$.

In Fig.\ 1, we plot the error measure $E$, defined in (\ref{E}), versus the coupling strength $\sigma$. Simulations are run for a total time duration $T=3000$, which is considerably larger than the typical time scale of an oscillation for an uncoupled Rossler oscillators, that is $2 \pi$; time averages are taken over the time interval $[2700,3000]$.

From the direct numerical integration of Eqs. (\ref{sync}) and (\ref{LD}) with $\bar{m}=0.2$, $\bar{p}=0$, $\delta \bar{a}=0$, and $\epsilon=\zeta=\eta=0$, we find that the master stability function converges to zero in the range $0.143 \ltsimeq \omega \ltsimeq 4.40$, which for our choice of the matrix $A$, corresponds to stability in the range $0.0113 \ltsimeq \sigma \ltsimeq 0.0511$. This range is delimited by the vertical dashed lines in Fig.\ 1, which shows good agreement with our computations of the full nonlinear system (\ref{ross}).
Figure 1 illustrates that the range of stability is affected neither by the presence of small deviations from the nominal couplings nor from small mismatches in the individual oscillators' parameters. This is because the eigenvalues $\lambda_i$ are indistinguishable from the eigenvalues $\ell_i$, for $i=2$ or $N$ to the degree of accuracy of the simulation shown in the figure. However, for $\sigma$ inside  the range of stability, the value attained by $E$ depends on the values of $\delta a_i$, $\delta m_i$, and $\delta p_i$.  Figure 2 shows $c_\epsilon$, $c_\zeta$, and $c_\eta$ versus $\omega$. With this information, Eq. (\ref{ccc}) provides an estimate of the master stability function for any tuple $(\omega,\epsilon,\zeta,\eta)$. We use Eq. (\ref{ccc}) along with the data plotted in Fig. 2 to calculate the master stability function $\mathcal{M}$. This is shown for comparison in Fig.\ 1, where the symbols $\times$ ($+$) are used to plot $\sum_{i=2}^N  \mathcal{M}(\omega_i,\epsilon_i,\zeta_i,\eta_i)^2$ for $\varsigma_a=10^{-4}$ and $\varsigma_m=\varsigma_p=0$ ($\varsigma_a=10^{-4}$ and $\varsigma_m=\varsigma_p=5 \times 10^{-4}$). Poorer agreement is observed for values of $\sigma$ slightly above the lower threshold for stability of $0.0113$ (not shown), which corresponds to a so-called \emph{bubbling} region, as further discusses below\footnote{
{
Numerical experiments 
performed by
replacing $\delta m_i$, $\delta p_j$, and $\delta A_{ij}$'s  with random numbers from the same distributions and $\lambda_1,...,\lambda_N$ and $W$  with the eigenvalues and eigenvectors of the original matrix $A^{NOM}$ show good agreement with the results  in Fig. 1.}
}.

\begin{figure}
\onefigure[width=8.5cm]{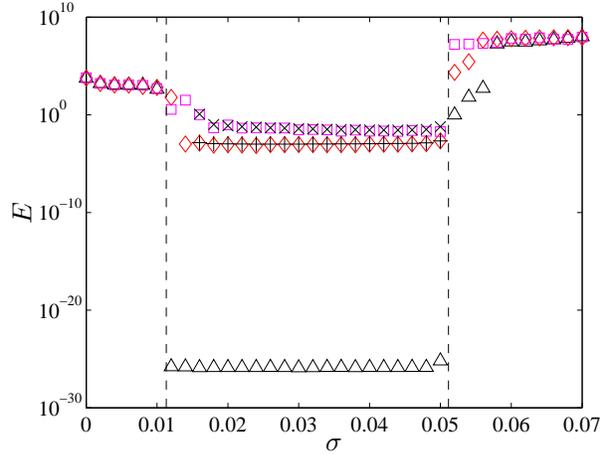}
\caption{Triangles, diamonds, and squares represent the error measure $E$ versus the coupling strength $\sigma$. 
Triangles are used for the case in which $\varsigma_a=\varsigma_m=\varsigma_p=0$. Diamonds are used for the case in which $\varsigma_a=10^{-4}$ and $\varsigma_m=\varsigma_p=0$. Squares are used for the case in which $\varsigma_a=10^{-4}$ and $\varsigma_m=\varsigma_p=5 \times 10^{-4}$. The vertical dashed lines delimit the range of stability predicted by the master stability function. The symbols $\times$ ($+$) refer to $\sum_{i=2}^N  \mathcal{M}(\omega_i,\epsilon_i,\zeta_i,\eta_i)^2$ for $\varsigma_a=10^{-4}$ and $\varsigma_m=\varsigma_p=0$ ($\varsigma_a=10^{-4}$ and $\varsigma_m=\varsigma_p=5 \times 10^{-4}$), computed using Eq. (\ref{ccc}).}
\label{fig.1}
\end{figure}

\begin{figure}
\onefigure[width=8.5cm]{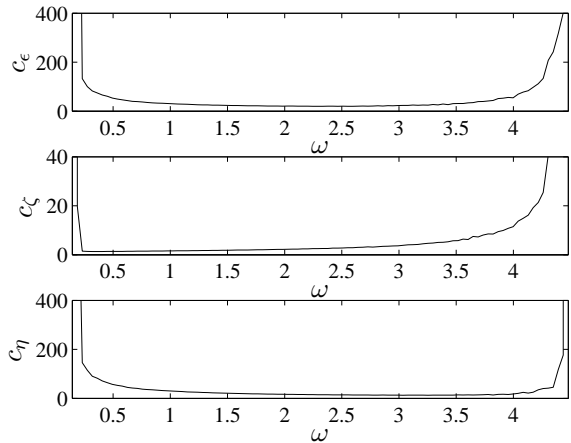}
\caption{$c_\epsilon$, $c_\zeta$, and $c_\eta$ versus $\omega$.}
\label{fig.1}
\end{figure}

{\bf Conclusions.} The master stability function analysis \cite{FujiYama83,Pe:Ca} provides a necessary and sufficient condition for linear stability of the synchronous solution for an arbitrary network of coupled identical systems. An extension of this approach  for networks of groups, where the dynamics of nodes within a group are the same but are different for nodes in distinct groups, is proposed in \cite{NSG}. In addition, a master stability function for networks in which each unit independently implements an adaptive strategy to maintain synchronization is presented in \cite{SAS}. The analysis of nearly identical coupled dynamical systems is considered in \cite{restr_bubbl,Su:Bo:Ni}. For this case, which is of practical relevance in experimental instances of synchronization and in biological systems, it is shown that a master stability function approach is applicable \cite{Su:Bo:Ni}.

In this letter, we have proposed a sensitivity analysis to address synchronization in the presence of
a broad range of deviations from nominal conditions. In particular, we have taken into consideration simultaneous small deviations in  the dynamics of individual units, the output functions of the individual units, and the coupling among the systems. We have shown that the master stability function formalism can be extended to this general scenario and that stability of the nearly-synchronous evolution depends on the eigenvalues  of an appropriately modified coupling matrix.
Our analysis is motivated by inherent practical challenges in implementing ideal conditions in experimental analysis of synchronization. For example, our approach can be directly applied to synchronization of nearly identical units whose interconnections yield to approximately zero-row-sum coupling matrix. 
In this case, the proposed master stability function can be used to estimate the conditions under which the nearly-synchronous evolution is stable and in case of stability, the approach can be used to quantify  the overall synchronization error.

Noise or small mismatches in the parameters of the individual systems can be responsible for the onset of \emph{bubbling} \cite{Ash1,restr_bubbl,SAS}, that is, rare intermittent large deviations from synchronization. We expect bubbling also to arise in the case of approximate satisfaction of the zero-row-sum condition; in this case, the master stability function, introduced in this letter, can be used to identify stable, unstable, and bubbling regions in the relevant parameter space, see for example \cite{SAS}.

\acknowledgments
F. Sorrentino would like to thank Ed Ott for insightful discussions. M. Porfiri was supported by the National Science Foundation under Grant No. CMMI-0745753.


%
%
%
%

\end{document}